# Cosmic walls and filaments formation in modified Chaplygin gas cosmology

**S. Karbasi** [(1)] and **H. Razmi**[(2)]

*Department of Physics, the University of Qom, 3716146611, Qom, I. R. Iran*

(1) s.karbasi@stu.qom.ac.ir

(2) razmi@qom.ac.ir & razmiha@hotmail.com

## Abstract

We want to study the perturbation growth of an initial seed of an ellipsoidal shape in Top-Hat collapse model of structure formation in the Modified Chaplygin gas cosmology. Considering reasonable values of the constants and the parameters of the model under study, it is shown that a very small deviation from spherical symmetry (ellipsoidal geometry) in the initial seed leads to a final highly non-spherical structure which can be considered as a candidate for justifying already known cosmological structures as cosmic walls and filaments.

## 1. Introduction

As the Universe expands, overdense regions break away from the Hubble flow, the regions collapse and form virialized objects, such as cluster of galaxy, filaments and walls. The spherical-collapse model is a basic well-known model in the theory of cosmic structure formation [1]. Collapsed haloes form from the initial fluctuation field, leading to structure formation; universe expands, sufficiently overdense regions expand until they reach a maximum size, afterward, they collapse under the action of their own gravity. This has been already enough studied using both analytical models and numerical simulations, in the standard cosmological model [2-7]. In this model, the structures are assumed to have grown gravitationally from small, initially Gaussian density fluctuations in the early universe [8]. According to the real observations of large scale structure reasoning on the cosmic web formation of the universe, the complicated ellipsoidal collapse model has been under consideration and study too [9-12]. As we know, cosmic walls, filaments, and galaxy clusters have non-spherical symmetry. There are a number of other necessary considerations and corrections in studying collapse models among them are asymmetry of initial seeds, the influence of cosmic expansion and neighboring heavy structures. It is well known that the protohalo mass distribution misses its spherical symmetry due to the tidal forces affecting on it during its evolution [13-17]. All the way, the final shape of the structures depends highly on the symmetry of their initial protohalos. Numerical simulations show that the virialized haloes can be considered as some triaxial objects [18-21]. Triaxial collapse models have been studied analytically with good results in the standard cosmology [22-23]. In this paper, we want to study ellipsoidal collapse model in Chaplygin gas (CG) cosmology. As we know, CG, as a strange fluid with an equation of state $p = -B/\rho$, $B > 0$, where $\rho$ and $p$ are the pressure and energy density respectively and B is a constant, plays the role of dark energy [24-25]. To be in good consistency with observational data, CG has been extended to the generalized Chaplygin gas (GCG) with the equation of state $p = -B/\rho^\alpha$ where $0 < \alpha < 1$ [26-29]. The modified Chaplygin gas (MCG) scenario, with the aim of considering both dark energy and dark matter, has been already considered with the equation of state $p = A\rho - B/\rho^\alpha$ [30-32]. MCG model not only explain the evolution of the universe from early stage to the present time [32-36], but also can be used in justifying the inflation and radiation dominated state of the universe [35-38]. In what follows, considering the well known 'Top-Hat' model in the large scale structure formation in GCG and MCG scenarios [39-41], we want to study ellipsoidal collapse model in MCG scenario.

## 2. Ellipsoidal collapse of Chaplygin gas

The ellipsoidal collapse model we study here is similar to the spherical one except that the shape of the initial seed is ellipsoidal. The background equations and all other necessary quantities and parameters (e.g. the density and the pressure) are as the same as what are in [39-41]:

$$\dot{\rho} = -3h(\rho + p) \quad , \quad \frac{\ddot{a}}{a} = -\frac{4\pi G}{3}\Sigma_j(\rho_j + 3p_j) \qquad (1),$$

except that, in the perturbed region, we consider the following relations:

$$\dot{\rho}_c = -3h(\rho_c + p_c) \quad , \quad \frac{\ddot{r}}{r} = -2\pi G \Sigma_j(\rho_{cj} + 3p_{cj})b_i \qquad (2),$$

where the index $c$ refers to the initial formed seed, $h \equiv \frac{\dot{r}}{r} = H + \frac{\theta}{3a}$, $\Theta \equiv \vec{\nabla}.\vec{v}$ ($\vec{v}$ is the peculiar velocity) is the expansion rate for the initial seed, and $b_i$'s are the coefficients usually introduced in the potential theory for homogeneous ellipsoids,

$$b_i = a_1 a_2 a_3 \int_0^\infty \frac{d\tau}{(x_i + \tau)\prod_{m=1}^3 (x_m^2 + 1)^{\frac{3}{2}}} \qquad (3)$$

where $(a_1, a_2, a_3)$ are semiaxes of the ellipsoid and the coefficients satisfy $\sum_{i=1}^3 b_i = 2$.

If two ones of the axes are equal, the integrals are reduced to elementary functions. Two interesting cases are an oblate spheroid with $a_1 = a_2 > a_3$

$$b_1 = b_2 = \frac{\sqrt{1-e^2}}{e^2}\left[\frac{\sin^{-1} e}{e} - \sqrt{1-e^2}\right] \qquad (4)$$

$$b_3 = \frac{2\sqrt{1-e^2}}{e^2}\left[\frac{1}{\sqrt{1-e^2}} - \frac{\sin^{-1} e}{e}\right] \qquad (5),$$

and a prolate spheroid with $a_1 = a_2 < a_3$

$$b_1 = b_2 = \frac{1-e^2}{e^2}\left[\frac{1}{\sqrt{1-e^2}} - \frac{1}{2e}\ln\left(\frac{1+e}{1-e}\right)\right] \qquad (6)$$

$$b_3 = \frac{2(1-e^2)}{e^2}\left[\frac{1}{2e}\ln\left(\frac{1+e}{1-e}\right) - 1\right] \qquad (7)$$

where the eccentricity is defined as

$$e = \sqrt{1 - \frac{a_3^2}{a_1^2}} \qquad (8).$$

Considering the density difference between the initial seed and the background as $\delta\rho$ and the density contrast as $\delta_j = \left(\frac{\delta\rho}{\rho}\right)_j$ where the index j refers to the Chaplygin gas or the baryons, the modified ellipsoidal evolution (the derivative with respect to the cosmic scale factor $a$) equation for $\delta_j$ is:

$$\delta'_j = -\frac{3}{a}(c_{effj}^2 - w_j)\delta_j - [1 + w_j + (1 + c_{effj}^2)\delta_j] \qquad (9)$$

where $c_{eff}^2 = \frac{\delta p}{\delta\rho}$ is the square of the effective sound speed and $w = \frac{p}{\rho}$ is the background state parameter.

The corresponding modified ellipsoidal evolution (the derivative with respect to the cosmic scale factor $a$) equation for $\theta$ is:

$$\theta' = -\frac{\theta}{a} - \frac{\theta^2}{3Ha^2}$$

$$-\frac{9}{4}\frac{H_0^2}{H}\left\{\Omega_{mcg}\left[(1+\delta_{mcg})b_i - \frac{2}{3}\right][\bar{c} + (1-\bar{c})a^{-3(1+A)(1+\alpha)}]^{\frac{1}{1+\alpha}}\right.$$

$$- \Omega_{mcg}\, 2A\, [\bar{c} + (1-\bar{c})a^{-3(1+A)(1+\alpha)}]^{\frac{1}{1+\alpha}} + \frac{2\bar{c}(1+A)\Omega_{mcg}}{[\bar{c} + (1-\bar{c})a^{-3(1+A)(1+\alpha)}]^{\frac{1}{1+\alpha}}}$$

$$+ \frac{\Omega_b}{a^3}\left[(1+\delta_b)b_i - \frac{2}{3}\right]$$

$$+ 3b_i\left[A\Omega_{mcg}[\bar{c} + (1-\bar{c})a^{-3(1+A)(1+\alpha)}]^{\frac{1}{1+\alpha}}(1+\delta_{mcg})\right.$$

$$\left.\left. - \frac{\bar{c}(1+A)\Omega_{mcg}}{[\bar{c} + (1-\bar{c})a^{-3(1+A)(1+\alpha)}]^{\frac{1}{1+\alpha}}(1+\delta_{mcg})^\alpha}\right]\right\} \qquad (10)$$

where $\bar{c} = \frac{B}{(1+A)\rho_0^{1+\alpha}}$ ($\rho_0$ is the density at the present time), and with the same already known notation and parameters introduced in [39-41].

## 3. The method and the result

In this section, the equations (9) and (10) in the ellipsoidal collapse model under consideration are integrated out using the same parameters, equations, numerical values of the redshift and the density parameters and the Hubble constant, and the same computational programming previously used in spherical collapse model in MCG cosmology [41] with the same initial conditions except that here the parameter $\bar{c}$ is fixed at $\bar{c} = 0.99$ ideally and the constant A is assumed to have an arbitrary value A= -0.02 with $\delta_{MCG}(z = 1000) = 3.5 \times 10^{-3}$ [40]. The constant $\alpha$ is assumed to have a value of $\alpha= 0.022$; this choice covers the accelerated expansion of the universe in MCG scenario. Assuming an initial small eccentricity of the value $e = 0.01$, the evolution of the rate of collapsed region, h, with respect to the redshift z, has been shown in figure 1 (2) for a prolate (an oblate) spheroid structure.

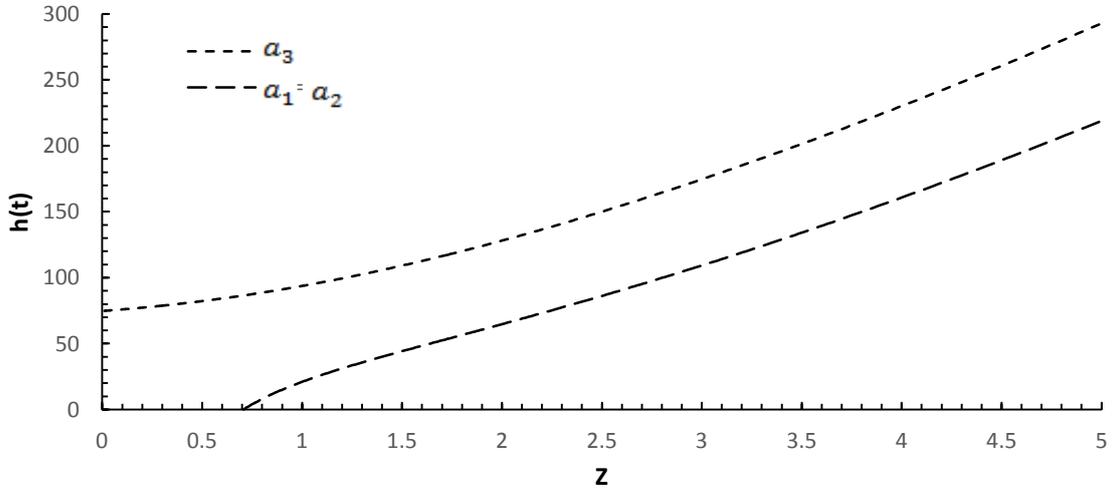

Fig. 1: Evolution of collapsed region expansion rate, h, with respect to the redshift z for a prolate spheroid structure ($a_1 = a_2 < a_3$).

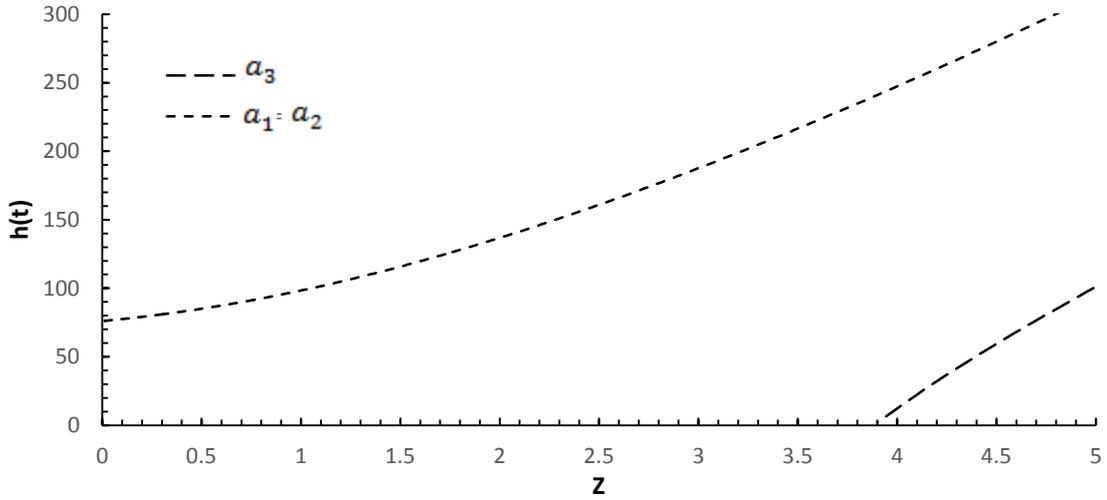

Fig. 2: Evolution of collapsed region expansion rate, h, with respect to the redshift z for an oblate spheroid structure $(a_1 = a_2 > a_3)$.

According to the definition of the parameter $h$ (the expansion rate of the seed), it is clear that when the value of this parameter approaches zero, the structure growth is stopped and collapses in itself. In this state, because of the contraction of the perturbed region, the structure under consideration is formed and its self gravitation makes rapid increase in its density; such a situation time is known as the time of the structure formation. If there is an initial small difference between the main axes of the ellipsoid, there are possible situations where this small difference can grow well after enough and suitable time of evolution so that the three dimensional structure transforms to the two dimensional walls and even to the one dimensional filaments. In the figures 1 and 2, one can clearly see the formation of non-spherical structures like filaments and/or walls. As is seen, the shorter axis began to collapse at earlier times relative to the longer one which grows with a rate near to the Hubble rate. Although the elongation rate of the greater axis is near to the Hubble rate, the structure growth still continues; this is what one expects for the filaments (walls) formation which continually become elongated (wider) and denser.

The square value of the adiabatic sound speed $c_s^2 = \frac{\partial p}{\partial \rho}$ has been shown in figure 3. As is seen, at lower than 1 values of the redshift, it is positive; this means that the adiabatic sound speed has a real value which guarantees the stability of the structure. In the early universe, at the higher values of z, the instability of the MCG fluid has a useful role in the structure formation [42].

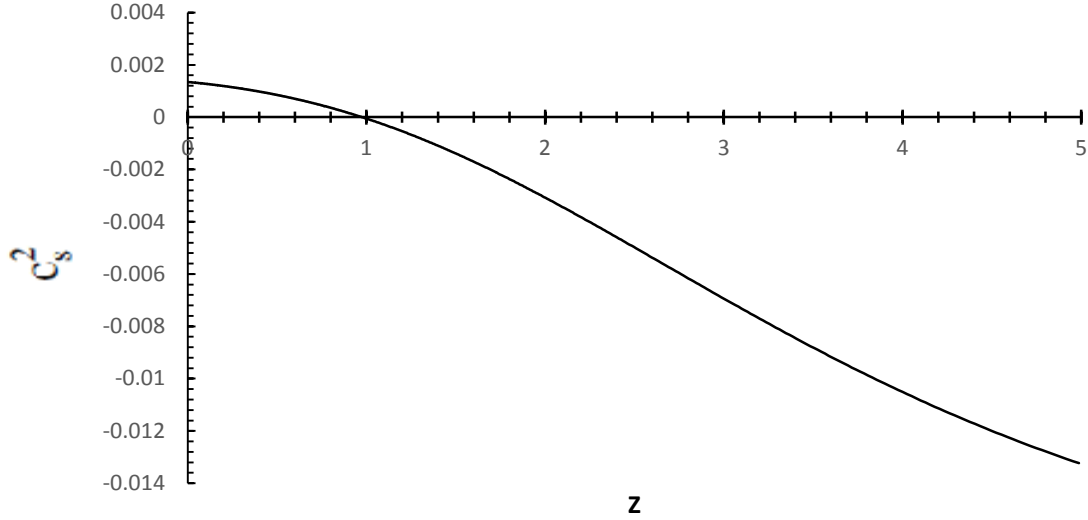

Fig.3: Evolution of the square value of the adiabatic sound speed $c_s^2$ with respect to the redshift

## 4. Conclusion

In this paper, we have studied perturbation growth in MCG-dominated universe in an ellipsoidal collapse model. As was seen, if the initial protohalo of the cosmic structure has a non-spherical (here ellipsoidal) geometry, then the final form of the structure can take elongated and/or wide shapes as filaments and/or walls. This has been shown in figures 1 and 2 schematically by analytical/numerical considerations with choosing particular values (which are consistent with the current theoretical and observational data) for the constant parameters appeared in the model. The stability of the system has been considered too. As is seen in figure 3, for $\alpha$=0.022, the system is stable; because the adiabatic sound speed has real value at the present time.